\documentclass[
aps,%
prd,
10pt,%
notitlepage,%
oneside,
nobibnotes,%
nofootinbib,%
superscriptaddress,%
noshowpacs,%
centertags]%
{revtex4-2}

\usepackage{epsf,amsmath,amsfonts,amssymb,amsbsy}
\usepackage[mathscr]{eucal}
\usepackage{hyperref}
\usepackage{graphicx}

\begin{document}
	
\title{Scalar Field Action under 4D Isotropic Cut-off and its Cosmological Impact}
	
\begin{abstract}
Specific action for isotropic fluctuations of scalar field is derived under the condition of 4D cut-off. It is implemented into the estimates of dark energy scale consistent with current cosmological data.
\end{abstract}

	\author{\firstname{Asya}~\surname{Aynbund}}
	\email{aynbund.asya@phystech.edu}
	\affiliation{Landau Phystech-School, Moscow Institute of Physics and Technology,
		Russia, 141701, Moscow Region, Dolgoprudny, Institutsky 9}
	
	\author{V.V.Kiselev}
	\email{kiselev.vv@phystech.edu; Valery.Kiselev@ihep.ru}
	
	\affiliation{Landau Phystech-School, Moscow Institute of Physics and Technology,
		Russia, 141701, Moscow Region, Dolgoprudny, Institutsky 9} 
	\affiliation{State
		Research Center of the Russian Federation  ``Institute for High Energy
		Physics'' of National Research Centre  ``Kurchatov Institute'',
		Russia, 142281, Moscow Region, Protvino, Nauki 1}
	
	\maketitle
\section{Introduction}	
The collaboration Dark Energy Spectroscopic Instrument (DESI) has provided analysis of cosmological data on baryonic acoustic oscillations \cite{DESI:2024uvr,DESI:2024lzq,DESI:2024mwx} that shows sensible deviations from the concordance $\Lambda$CDM model (abbreviated from ``cosmological constant $\Lambda$-term and Cold Dark Matter'') in favour of Dark Energy (DE) approximated by the pressure to energy density ratio 
$$
	w_\textsc{de}=\frac{p_\textsc{de}}{\rho_\textsc{de}}=w_0+w_a\cdot (1-a),
$$
in vicinity of current scale factor $a(t_0)=1$. Although estimates in the appropriate $w_0w_aCDM$ model \cite{DESI:2024uvr}
\begin{equation}
\label{desi-1}
	w_0=-0.827\pm 0.063,\qquad w_a=-0.75^{+0.29}_{-0.25}
\end{equation}
are still not in 5$\sigma$ contradiction to $\Lambda$CDM. In both models the energy density scale is in question since it is about $10^{-3}$ eV much less than any reasonable evaluation in physics of Universe cosmology \cite{Weinberg:1988cp,Zeldovich:1968ehl,Baumann:2018muz}. In this respect one can consider the stress-energy tensor $T_{\mu\nu}^\mathrm{cl.}$ entering the Einstein equations of general relativity in a way setting this tensor in terms of a density matrix $\hat \rho$ for non-stationary quantum states of fields as
\begin{equation}
\label{stress-1}
	T_{\mu\nu}^\mathrm{cl.}=\mathrm{tr}\big(\hat \rho\, \hat T_{\mu\nu}\big).
\end{equation}
There are two options for selecting the contribution of minimal energy states in (\ref{stress-1}). 

The first option is given by an invariant vacuum common for all fields, while the invariance implies that the vacuum contribution is separable in contrast to all other states entangled to give the following expression:
\begin{equation}
\label{stress-vac+}
	\mathrm{tr}\big(\hat \rho\, \hat T_{\mu\nu}\big)_\mathrm{inv.}=
	W_\mathrm{vac}\cdot\langle\mathrm{vac}|\hat T_{\mu\nu}|\mathrm{vac}\rangle+
	\mathrm{tr}_\mathrm{non-vac}\big(\hat \rho\, \hat T_{\mu\nu}\big)=
	W_\mathrm{vac}\cdot\rho_\mathrm{vac}^\mathrm{bare}\,g_{\mu\nu}+T_{\mu\nu}\big[\Phi_\mathrm{cl.}\big],
\end{equation}
where $W_\mathrm{vac}$ is the constant probability of vacuum in the density matrix, $\rho_\mathrm{vac}^\mathrm{bare}$ is the constant energy density of invariant vacuum, $g_{\mu\nu}$ is the metric, and $T_{\mu\nu}\big[\Phi_\mathrm{cl.}\big]$ is the stress-energy tensor of classic fields due to entanglement of quantum fields in the non-stationary University state. In the current paper we consider the origin of invariant vacuum due to isotropic 4-dimensional fluctuations in a coherent non-stationary state of scalar inflaton-field that substantiates the natural value of vacuum probability
\begin{equation}
\label{W-vac}
	W_\mathrm{vac}=\mathrm e^{-n_\mathrm{eff}}
\end{equation} 
with $n_\mathrm{eff}=\langle\mathrm{vac}|\hat n|\mathrm{vac}\rangle$ being the average number of quanta of invariant scalar field in the coherent state. At $\rho_\mathrm{vac}^\mathrm{bare}\sim \Lambda^4$ set by the inflation scale $\Lambda\sim 10^{16}$ GeV, with the reduced Planck mass $\tilde m_\mathrm{Pl}$ related with the Newton constant $G$,
\begin{equation}
\label{int-3}
	\tilde m_\mathrm{Pl}=(8\pi G)^{-1}
\end{equation} 
the estimate $n_\mathrm{eff}\sim \tilde m_\mathrm{Pl}/\Lambda$ gives $n_\mathrm{eff}\sim 250$ 
and results in inspiring match to the observed scale of energy density associated to the cosmological constant  \cite{Balitsky:2016gwm,Balitsky:2014epa,Kiselev:2013zia,Aynbund:2024lby}.

This study is inherently related with the second option: the entangled minimal energy state marked by the same symbol $|\mathrm{vac}\rangle$, is not separable, hence, the entanglement results in a non-invariant stress-energy tensor with diagonal components of energy density $\rho_\textsc{de}$ and pressure $p_\textsc{de}$
\begin{equation}
\label{stress-de}
	\rho_\textsc{de}=W_\textsc{de}\cdot \langle\mathrm{vac}|\hat T_{00}|\mathrm{vac}\rangle\cdot g^{00}=
	W_\textsc{de}\cdot \rho_\mathrm{vac}^\mathrm{bare},\qquad
	p_\textsc{de}=w_\textsc{de}\cdot \rho_\textsc{de},
\end{equation}
where the dark energy components would satisfy the conservation law in the expanding Universe
\begin{equation}
\label{de-conserve}
	\dot \rho_\textsc{de}+3H (\rho_\textsc{de}+p_\textsc{de})=0,
\end{equation}
under $H\equiv {\dot a}/{a}$ being the Hubble parameter ($\dot a = \mathrm d a / \mathrm d t$). This equation is equivalent to
\begin{equation}
\label{de-conserve-2}
	\dot W_\textsc{de}\cdot \rho_\mathrm{vac}^\mathrm{bare}+3H (1+w_\textsc{de})
	\,W_\textsc{de}\cdot \rho_\mathrm{vac}^\mathrm{bare} =0.
\end{equation}
So, introducing the e-folding in terms of scale factor $a(t)$
$$
	N\stackrel{\mathrm{def}}{=}\ln a(t),	
$$
we note that the time derivative equals
\begin{equation}
\label{derive}
	\dot W_\textsc{de}=\frac{\mathrm d  W_\textsc{de}}{\mathrm d N}\cdot \frac{\mathrm d N}{\mathrm d a}
 \cdot \frac{\mathrm d a}{\mathrm d t}
 =\frac{\mathrm d  W_\textsc{de}}{\mathrm d N}\cdot \frac{\dot a}{a}=
 \frac{\mathrm d  W_\textsc{de}}{\mathrm d N}\cdot H\equiv W_\textsc{de}^\prime\cdot H .
 \end{equation}
Therefore, eq. (\ref{de-conserve-2}) takes the form
\begin{equation}
\label{de-conserve-3}
	w_\textsc{de}=-1-\frac13\,\big(\ln W_\textsc{de}\big)^\prime.
\end{equation}
In vicinity of $N= 0$ one can introduce the effective number of quanta
\begin{equation}
\label{n-eff}
	n_\mathrm{eff}(N)\approx n_\mathrm{eff}+n^\prime\cdot N+\frac12\,n^{\prime\prime}\cdot N^2,
\end{equation}
where $n_\mathrm{eff}=n_\mathrm{eff}(0)$, $n^\prime$ and $n^{\prime\prime}$ are constant, while
\begin{equation}
\label{W-de}
	W_\textsc{de}=\mathrm e^{-n_\mathrm{eff}(N)}.
\end{equation}
 Then,
\begin{equation}
\label{de-conserve-4}
	w_\textsc{de}=-1+\frac13\,\big(n^\prime+n^{\prime\prime}\cdot N\big)\approx
	-1+\frac13\,\big(n^\prime-n^{\prime\prime}\cdot(1-a)\big).
\end{equation}
Thus, 
$$
	w_0=-1+\frac13\,n^\prime,\qquad w_a=-\frac13\,n^{\prime\prime}.
$$
Empirical data  \cite{DESI:2024uvr} correspond to approximate constraints
\begin{equation}
\label{ww-1}
	0.3 < n^\prime <0.75,\qquad 1.7 < n^{\prime\prime} < 3,
\end{equation}
The model estimations of both $n^{\prime}$ and $n^{\prime\prime}$ could be derived from some Lindblad-like equations 
\cite{Lindblad:1975ef} for the density matrix of the scalar field vacuum-state entangled to environment. 

In this context it is important to note that both described options are based on the same mechanism for suppressing the bare energy density of scalar field vacuum. 

In the present paper we consider a role of cut-off to find the action for the scalar field if the fluctuations are 4D isotropic. The cut-off in the local field theory itself is inevitable if we assume that the ultimate theory is based on non-local superstrings \cite{Halverson:2018vbo,Kinney:2009vz}. On the other hand, one could presume that a quantum gravity has to  somehow involve a quantum cells of space-time, and those cells have to be characterised by an invariant scale of minimal size associated with a maximal energy scale or a cut-off, too. Therefore, the problem of 4D isotropic action is actual, indeed. 

So, calculating a contribution into a density of vacuum energy by zero-point modes for a scalar field involves a 4D isotropic cut-off $\Lambda$ in a momentum space in order to make this contribution to be  finite and to get invariant stress-energy tensor $T^\mu_\nu=\rho_\mathrm{vac}\,\delta^\mu_\nu$ for the vacuum with the energy density $\rho_\mathrm{vac}$.
\section{Substantiating the action}
The action of real scalar field $\phi(x)$ takes the form 
\begin{equation}
\label{act1}
	S=\int\mathrm{d}^4x\,\frac12\big(\phi(x)\,(-\partial_\mu\partial^\mu\phi(x))-m^2\phi^2(x)\big),
\end{equation}
where the integration takes place over 4D volume $V_\mathrm{4D}\to \infty$, while the dimension of scalar field equals the dimension of mass: $[\phi(x)]=[\Lambda]$. Under the Fourier transform
\begin{equation}
\label{act2}
	\phi(x)=\int\frac{\mathrm{d}^4p}{(2\pi)^4}\,\mathrm{e}^{-\mathrm{i}p_\mu x^\mu}\,\Phi[p]
\end{equation}
the action reads off
\begin{equation}
\label{act3}
	S=\int\frac{\mathrm{d}^4p}{(2\pi)^4}\,\frac12\,\Phi^*[p](p^2-m^2)\Phi[p],
\end{equation}
and the dimension of $\Phi[p]$ equals $[\Phi]=[\phi(x)]/[\Lambda]^4$.

In order to get the 4D isotropic solutions with a cut-off alike $\Lambda$ we perform Wick rotation $p_0=\mathrm{i}p_4$ with $p^2=-p_{\textsc{e}}^2$ and $\mathrm{d}^4p=\mathrm{i}\mathrm{d}\Omega_3\,p_{\textsc{e}}^3\mathrm{d}p_{\textsc{e}}$, where the solid angle of 3D sphere is equal to 
$$
	\int\mathrm{d}\Omega_3=\int\limits_{-1}^{+1}\mathrm{d}\cos\theta_1
	\int\limits_{-1}^{+1} \sin\theta_1\mathrm{d}\cos\theta_2
	\int\limits_0^{2\pi}\mathrm{d}\varphi=2\pi^2.
$$
For 4D isotropic solutions $\Phi[p]\mapsto \Phi(p_{\textsc{e}})$ and the action is transformed into the expression
\begin{equation}
\label{act4}
	S_{\textrm{4D}}=\mathrm{i}\frac{1}{(2\pi)^3}\,\int \mathrm{d}\Omega_3\cdot
	\int p_{\textsc{e}}^3\,\frac{\mathrm{d}p_{\textsc{e}}}{2\pi}\,\frac12 \Phi^*(p_{\textsc{e}})
	(-p_{\textsc{e}}^2-m^2)\Phi(p_{\textsc{e}})=\frac{\mathrm{i}}{4\pi}\,
	\langle p_{\textsc{e}}^3\rangle_\Lambda  
	\int \frac{\mathrm{d}p_{\textsc{e}}}{2\pi}\,\frac12 \Phi^*(p_{\textsc{e}})
	(-p_{\textsc{e}}^2-m^2)\Phi(p_{\textsc{e}}),
\end{equation}
taking into account the fact of imposing a cut-off of the order of $\Lambda$ that makes the integration to be finite and allowing for averaging $\langle p_{\textsc{e}}^3\rangle_\Lambda \sim\Lambda^3$. The back Wick rotation $\mathrm{i}p_{\textsc{e}}=k$ provides 
\begin{equation}
\label{act5}
	S_{\textrm{4D}}=\frac{1}{4\pi}\,
	\langle p_{\textsc{e}}^3\rangle_\Lambda  
	\int \frac{\mathrm{d}k}{2\pi}\,\frac12 \Phi^*(k)
	(k^2-m^2)\Phi(k), 
\end{equation}
while under 
$$
	\Phi(k)=\frac{1}{\Lambda^3}\int\mathrm{d}\tau\,\mathrm{e}^{\mathrm{i}\tau k}\phi(\tau)
$$
we arrive to 
\begin{equation}
\label{act6}
	S_{\textrm{4D}}=\frac{1}{4\pi}\,
	\frac{\langle p_{\textsc{e}}^3\rangle_\Lambda}{\Lambda^6}  
	\int \mathrm{d}\tau\,\frac12 \phi(\tau)
	\left(-\frac{\mathrm{d}^2}{\mathrm{d} \tau^2}-m^2\right)\phi(\tau).
\end{equation}
Writing down eqs. (\ref{act5}) and (\ref{act6}) we purposely repeat the same symbols $\Phi$ and $\phi$ despite the substitutions in the arguments.

So, introducing a new cut-off scale $\Lambda$ of the same order of magnitude by 
$$
	\frac{1}{4\pi}\,\frac{\langle p_{\textsc{e}}^3\rangle_\Lambda}{\Lambda^6}\mapsto
	\frac{1}{\Lambda^3},
$$
we substantiate the action under 4D isotropic cut-off
\begin{equation}
\label{act7}
	S_{\textrm{4D}}=\frac{1}{\Lambda^3}\int \mathrm{d}\tau\,\frac12\,(\dot\phi^2-m^2\phi^2),
\end{equation}
where $\dot\phi=\mathrm{d}\phi/\mathrm{d}\tau$, and we have used the integration by parts.

\section{Deductive statements}
At first, the action (\ref{act7}) looks like the action of scalar field in the finite volume, though one has to note the invariant constant 3D volume $V_\mathrm{3D}=1/\Lambda^3$, while the variable $\tau$ is also Lorentz invariant. Therefore, the 4D isotropic cut-off for the invariant solutions of scalar field can be equivalently described by the action in a finite invariant volume. 

Second, the action obtained is the action of harmonic oscillator. Hence, the quantum description is ordinary and finite. 

Third, formally one can get the action (\ref{act7}) from the action of scalar field (\ref{act1}) under the following operational substitutions:
$$
	\phi(x)\mapsto \phi(\tau)\quad\mbox{and}\qquad
	\int\mathrm{d}^4x\ldots\partial_\mu\ldots\partial_\nu\ldots\;\mapsto\;
	\frac{1}{\Lambda^3}\int \mathrm{d}\tau\,\ldots\frac14 \,g_{\mu\nu}
	\frac{\mathrm{d}}{\mathrm{d}\tau}\ldots\frac{\mathrm{d}}{\mathrm{d}\tau}\ldots
$$
as it was explored in \cite{Aynbund:2024lby} in view of the invariant description of isotropic fluctuations for the inflaton field.

The most spectacular feature of action transformation is the consequence for the stress-energy tensor since one gets the following expression for 4D-isotropic solutions subject to the cut-off: 
\begin{multline}
\label{act8}\textstyle
	T_{\mu\nu}=(\partial_\mu\phi)(\partial_\nu\phi)-\frac12\,g_{\mu\nu}
	\big((\partial_\lambda\phi)^2-m^2\phi^2\big)\;\mapsto\; \\ {} \textstyle\mapsto\;
	T_{\mu\nu}^\Lambda=\frac14\,g_{\mu\nu}\dot\phi^2-\frac12\,g_{\mu\nu}
	\big(\dot\phi^2-m^2\phi^2\big)= 
	- \frac14\,g_{\mu\nu}\dot\phi^2+\frac12\,g_{\mu\nu} 	m^2\phi^2.\quad
\end{multline}

For the oscillator at $m\neq 0$ the contribution into the energy density by the positive potential term always exceeds the negative kinetic term in (\ref{act8}).  Thus, the covariant structure of space-time in the finite volume results in the vacuum-like stress-energy tensor at least for the basic state with no quanta of oscillator and the bare value of energy density  (see details in \cite{Aynbund:2024lby})
\begin{equation}
\label{cov-6}
	\rho_\mathrm{vac}^\mathrm{bare}=\langle \mbox{vac} |T_{00}^\Lambda |\mbox{vac}\rangle =\frac18\,m\Lambda^3>0.
\end{equation}
Further numerical estimates in cosmology with the inflaton field repeat values announced in Introduction: the scheme is consistent to fit the  empirical scale of vacuum energy density.

Another questions is what does happen if $m=0$? We certainly conclude that the 4D-isotropic massless field with the cut-off generates the anti-de Sitter like stress-energy tensor 
\begin{equation}
\label{phan}
	T_{\mu\nu}^\Lambda=- \frac14\,g_{\mu\nu}\dot\phi^2
\end{equation}
with a negative density of energy at $\dot\phi=\mbox{const.}$, setting the constraint which satisfies the equation of motion.  The phantom-lke form of (\ref{phan}) physically signalise that 4D isotropic fluctuations are unstable and cannot produce separable density matrix, and such a quantum state has to be entangled with environment so that the state has to decay. The decay time has to exceed the minimal admissible time $\tau> 1/\Lambda$, wherein a factor $C$ in $\tau=C/\Lambda$ may be much grater than 1, too. 

The entanglement of phantom scalar would result in the phantom spatially isotropic massless field depending on time $\phi=\phi(t)$. Such a field is responsible for wormholes found by Ellis and Bronnikov \cite{Ellis:1973yv,Bronnikov:1973fh}. 

The same field with a stiff matter equation of state ($w=+1$) generates spatially isotropic constant Einstein tensor $G_{\mu\nu}=R_{\mu\nu}-R \,g_{\mu\nu}/2$ that produces anisotropic evolution of universe with expanding and contracting spatially flat subspaces \cite{Kiselev:2018iru}. This effect could explain the dynamical formation of 3 ordinary expanding dimensions and  extra contracting dimensions of space so that extra contracting dimensions could to be compactified into small volumes because of the contraction. 

Anyway such exotic effects cannot be long in time since the entanglement of phantom field has to cause a finite interval of operation in deal.

\section{Conclusion}
We have derived the action of scalar field incorporating the isotropic 4D cut-off. Such the action is important in empirical cosmology as concerns for the scale of cosmological constant and recent detection of effects in the dark energy equation of state. The entanglement of scalar vacuum state requires further studies in terms of Lindblad equation \cite{Lindblad:1975ef} specified for the cosmology.

To the same moment the case of massless field allows us to consider new aspects of unstable phantom field in exotic wormholes and in primordial steps of universe evolution to dynamically segregate compactified extra dimensions.

\bibliography{bib-4Dcut}	
\end{document}